\documentclass[english]{article}
\usepackage[T1]{fontenc}
\usepackage[latin1]{inputenc}
\usepackage{amsmath}
\usepackage{graphicx}
\usepackage{amssymb}
\IfFileExists{url.sty}{\usepackage{url}}
                      {\newcommand{\url}{\texttt}}

\makeatletter

\newcommand{\noun}[1]{\textsc{#1}}

\newcommand{\lyxaddress}[1]{
\par {\raggedright #1
\vspace{1.4em}
\noindent\par}
}

\usepackage{babel}
\makeatother
\begin{document}

\title{The $\beta$-neurexin/neuroligin-1 interneuronal intrasynaptic adhesion
is essential for quantum brain dynamics}

\author{Danko Dimchev Georgiev}

\maketitle

\lyxaddress{Laboratory of Molecular Pharmacology, Faculty of Pharmaceutical Sciences,
Kanazawa University Graduate School of Natural Science and Technology,
Kakuma-machi, Kanazawa, Ishikawa 920-1192, JAPAN\\
E-mail: \url{danko@p.kanazawa-u.ac.jp}}

\begin{abstract}
There are many blank areas in understanding the brain dynamics and
especially how it gives rise to consciousness. Quantum mechanics is
believed to be capable of explaining the enigma of conscious experience,
however till now there is not good enough model considering both the
data from clinical neurology and having some explanatory power. In
this paper is presented a novel model in defence of macroscopic quantum
events within and between neural cells. The synaptic $\beta$-neurexin/neuroligin-1
adhesive protein complex is claimed to be not just the core of the
excitatory glutamatergic CNS synapse, instead it is a device mediating
entanglement between the cytoskeletons of the cortical neurons. Thus
the macroscopic coherent quantum state can extend throughout large
brain cortical areas and the subsequent collapse of the wavefunction
could affect simultaneously the subneuronal events in millions of
neurons. The neuroligin-1/$\beta$-neurexin/synaptotagmin-1 complex
also controls the process of exocytosis and provides an interesting
and simple mechanism for retrograde signalling during learning-dependent
changes in synaptic connectivity. A brief outlook of the molecular
machinery driving neuromediator release through exocytosis is provided
with particular emphasis on the possibility for vibrationally-assisted
tunneling.
\end{abstract}
\begin{verse}
\noun{Keywords}: quantum brain dynamics, Bose-Einstein condensation,
macroscopic quantum coherence, neuron, synapse, $\beta$-neurexin,
neuroligin-1, synaptotagmin-1
\end{verse}

\section{Subneuronal macroscopic quantum coherence}

There are a couple of models trying to resolve the enigmatic feature
of consciousness. The most popular is the Hameroff-Penrose Orch OR
model (Hameroff and Penrose, 1998) supposing that microtubule network
within the neurons and glial cells acts like a quantum computer. The
tubulins are in superposition and the collapse of the wave function
is driven by the quantum gravity. A string theory model is developed
by Nanopoulos and Mavromatos (Nanopoulos, 1995; Nanopoulos and Mavromatos,
1995) that is further refined into QED-cavity model (Mavromatos, 2000;
Mavromatos et al., 2002) suggesting dissipationless energy transfer
and biological quantum teleportation.

The macroscopic quantum coherence is defined as a quantum state governed
by a macroscopic wavefunction, which is shared by multiple particles.
This typically involves the spaciotemporal organization of the multiparticle
system and is closely related to Bose-Einstein condensation. Examples
of quantum coherence in many particle macroscopic systems include
superfluidity, superconductivity, and the laser. Of these three paradigm
systems, the former two (superfluidity and superconductivity) are
basically equilibrium systems, whereas the laser is our first example
of an open system, which achieves coherence by energetic pumping -
this latter idea is of the greatest importance for understanding the
general implications of coherence. The laser functions at room temperature
and is typical nonequilibrium possibility for coherence to exist and
endure at macroscopic and thermally challenging scales.

It is expected that observable quantum effects in biological matter
be strongly suppressed mainly due to the macroscopic nature of most
biological entities, as well as the fact that such systems live at
near room temperature. These conditions normally result in a very
fast collapse of the pertinent wave functions to one of the allowed
classical states. The brain operates at 310K and deviations in brain
temperature in either direction are not well tolerated for consciousness.
This temperature is quite toasty compared to the extreme cold needed
for quantum technological devices which operate near absolute zero.
In technology, the extreme cold serves to prevent thermal excitations,
which could disrupt the quantum state. However proposals for biological
quantum states suggest that biological heat is used to pump coherent
excitations. In other words biomolecular systems may have evolved
to utilize thermal energy to drive coherence. The assumption/prediction
by quantum advocates is that biological systems (at least those with
crystal lattice structures) have evolved techniques to funnel thermal
energy to coherent vibrations conducive to quantum coherence, and/or
to insulate quantum states through gelation or plasma phase screens
(Hagan et al., 2002).

The neural cytoplasm exists in different phases of liquid sol and
solid gel. Transition between sol and gel phases depends on actin
polymerization. Triggered by changes in Ca$^{2+}$ concentration,
actin copolymerizes with different types of actin cross-linking proteins
to form dense meshwork of microfilaments and various types of gels
which encompass microtubules and organelles. The particular type of
actin cross-linkers determines characteristics of the actin gels.
Gels depolymerize back to liquid phase by Ca$^{2+}$ activation of
gelsolin protein, which severs actin. Actin repolymerizes into gel
when Ca$^{2+}$ concentration is reduced. Actin gel, ordered water
jello phases alternate with phases of liquid cytosol. Exchange of
Ca$^{2+}$ between actin, microtubules and microtubule-bound calmodulin
can mediate such cycles. The transition between the alternating phases
of solution and gelation in cytosol depends on the polymerization
of actin, and the particular character of the actin gel in turn depends
on actin cross-linking. Of the various cross-linker related types
of gels, some are viscoelastic, but others (e.g. those induced by
the actin cross-linker avidin) can be deformed by an applied force
without response. Cycles of actin gelation can be rapid, and in neurons,
have been shown to correlate with the release of neurotransmitter
vesicles from presynaptic axon terminals. In dendritic spines, whose
synaptic efficacy mediates learning, rapid actin gelation and motility
mediate synaptic function, and are sensitive to anesthetics. Therefore
the actin gelation within neurons might favour water ordering and
quantum coherence at body temperature.

Even in the liquid phase, water within cells is not truly liquid and
random. Pioneering work by Clegg (1984) have shown that water within
cells is to a large extent ordered, and plays the role of an active
component rather than inert background solvent. Neutron diffraction
studies indicate several layers of ordered water on such surfaces,
with several additional layers of partially ordered water. Thus the
actin meshwork that encompasses the microtubules orders the water
molecules in the vicinity and might shield the quantum coherence/entanglement
between microtubule tubulins.

\section{On the dynamically ordered structure of water}

Within the framework of quantum field theory Jibu and colleagues have
described the dynamically ordered structure of water in the brain
(Jibu et al., 1994; 1996; Jibu and Yasue, 1997). In the discussion
following we briefly review their model. First, let us denote the
spatial region immediately adjacent to the cytoskeletal proteins by
$\mathbb{V}$ and introduce Cartesian system of coordinates $\mathcal{O}(x,y,z)$
so that any point of the region can be labelled by giving its coordinates
$r=(x,y,z)$. This volume is filled with water molecules and ions
(the number of ions is relatively small, less than $1\%$ of the total
number). Potassium ions (K$^{+}$) have radius equal to the radius
of the water molecules, so potassium ions can be mixed with water
in the dynamically ordered state. Sodium ions (Na$^{+}$) and calcium
ions (Ca$^{2+}$) have radii smaller than the radius of the water
molecules so they do not disturb the dynamically ordered structure
of water. Chloride (Cl$^{-}$) and hydrogencarbonate (HCO$_{3}^{-}$)
ions however have larger radii than the radius of water molecules.
If there are chloride or hydrogencarbonate ions in the region, then
the system of the radiation field and water molecules will suffer
from dynamical disorder and so the dynamically ordered structure of
water manifests defects. In normal physiological state of neurons
the chloride ions have very low concentrations inside neurons (3 mM)
compared to the extraneuronal space (110 mM). Intraneuronal Cl$^{-}$
concentration increases under GABA$_{\text{A}}$ receptor activation
by general anesthetics, therefore interfering with the dynamical ordering
of water molecules might explain loss of consciousness during general
anesthesia.

Further, we fix the total number $N$ of water molecules in the region
$\mathbb{V}$. If we look at the j$^{\text{th}}$ water molecule ($j$
running from $1$ to $N$ denotes fictitious number labeling the $N$
water molecules in question), its position will be given by coordinates
$r^{j}=(x^{j},y^{j},z^{j})$. From a physical point of view, a water
molecule has a constant electric dipole moment. The average moment
of inertia and electric dipole moment of a water molecule are respectively
estimated to be $I=2m_{p}d^{2}$ with $d=0.82\textrm{\AA}$ and $\mu=2e_{p}P$
with $P=0.2\textrm{\AA}$. Here $m_{p}$ denotes the proton mass and
$e_{p}$ the proton charge. Due to the electric dipole moment $\mu$
the water molecule interacts strongly with the radiation field in
the spatial region $\mathbb{V}$. Although the water molecules have
many energy eigenstates and so can exchange energy with the radiation
field in many different values, we restrict the discussion to the
case in which only the two principle eigenstates can take part in
the energy exchange. These are taken to be low lying states such that
either the probability of transition between two other eigenstates
is low relative to that between the two principal eigenstates or the
equilibrium populations of the other levels become sufficiently small
to allow them to be ignored. This coincides with the conventional
two-level approximation in describing energy exchange between atoms
and the radiation field in laser theory. Then one sees immediately
that the quantum dynamics of the j$^{\text{th}}$ water molecule can
be described by a fictitious spin variable $s^{j}=\frac{1}{2}\sigma$
in energy spin space, where $\sigma=(\sigma_{x},\sigma_{y},\sigma_{z})$
and the $\sigma_{i}$'s are the Pauli spin matrices denoting the three
components of the angular momentum for spin $\frac{1}{2}$.

Let $\epsilon$ be the energy difference between the two principal
energy eigenstates of the water molecule. Its actual value is $\epsilon\approx24.8$
meV. Then the Hamiltonian governing the quantum dynamics of the j$^{\text{th}}$
water molecule is given by $\epsilon s_{z}^{j}$ and the total Hamiltonian
for $N$ water molecules becomes:

\begin{equation}
H_{WM}=\epsilon\sum_{j=1}^{N}s_{z}^{j}\end{equation}

The two eigenvalues of this Hamiltonian are $-\frac{1}{2}\epsilon$
and $\frac{1}{2}\epsilon$ reflecting the fact that only the two principal
energy eigenstates with energy difference $\epsilon$ have been taken
into account.

Now let us consider the radiation in the spatial region $\mathbb{V}$
from the point of view of quantum field theory. It is convenient to
describe the radiation field in terms of its effect on an electric
field operator $E=E(r,t)$. We assume for simplicity that electric
field is linearly polarized, obtaining $E=eE$, where $e$ is a constant
vector of unit length pointing in the direction of linear polarization.
Then, the radiation field in question comes to be described by a scalar
electric field $E=E(r,t)$ governed by the usual Hamiltonian:

\begin{equation}
H_{EM}=\frac{1}{2}\int_{V}E^{2}d^{3}r\end{equation}

Next, we introduce the interaction between the radiation field and
the totality of water molecules by which they can exchange energy
in terms of the creation and annihilation of photons, that is energy
quanta, of the radiation field. The electric field operator can be
divided into positive and negative frequency parts $E=E^{+}+E^{-}$.
Then the interaction Hamiltonian of the radiation field and the totality
of water molecules becomes:

\begin{equation}
H_{I}=-\mu\sum_{j=1}^{N}\left[E^{-}(r^{j},t)s_{-}^{j}+s_{+}^{j}E^{+}(r^{j},t)\right]\end{equation}

where $s_{\pm}^{j}=s_{x}^{j}\pm s_{y}^{j}$ are ladder operators in
energy spin space. The total Hamiltonian governing the quantum dynamics
of the radiation field, the electric dipoles of the water molecules,
and their interaction is given by

\begin{equation}
H=H_{WM}+H_{EM}+H_{I}\end{equation}

Since the region $\mathbb{V}$ maybe considered as a cavity for the
electromagnetic wave, it is convenient to introduce the normal mode
expansion of the electric field operator $E=E^{+}+E^{-}$, obtaining

\begin{equation}
E^{\pm}(r,t)=\sum_{k}E_{k}^{\pm}(t)\exp\left[\pm\imath(\kappa\cdot r-\omega_{k}t)\right]\end{equation}

Here, $\omega_{k}$ denotes the proper angular frequency of the normal
mode with wave vector $\kappa$. We are mainly interested in the ordered
collective behaviour among the water molecules and the radiation field
in the region $\mathbb{V}$, i.e. the vicinity of the cytoskeletal
proteins. Let us introduce therefore collective dynamical variables
$S_{k}^{\pm}(t)$ for the quantized electromagnetic field given by

\begin{equation}
S_{k}^{\pm}(t)\equiv\sum_{j=1}^{N}s_{\pm}^{j}(t)\exp\left[\pm\imath(\kappa\cdot r^{j}-\omega_{k}t)\right]\end{equation}

and the collective dynamical variable for the water molecules given
by

\begin{equation}
S\equiv\sum_{j=1}^{N}s_{z}^{j}\end{equation}

Then, the total Hamiltonian $H$ becomes

\begin{equation}
H=H_{EM}+\epsilon S-\mu\sum_{k}\left[E_{k}^{-}S_{k}^{-}+E_{k}^{+}S_{k}^{+}\right]\end{equation}

This total Hamiltonian for the system of $N$ water molecules and
the radiation field in the region $\mathbb{V}$ around the cytoskeletal
protein network is essentially of the same form as Dicke's Hamiltonian
for the laser system. Therefore it might be expected that water should
manifest laser-like coherent optical activity, that is, act as a water
laser under certain biologically realizable conditions.

\section{Synaptogenesis in the mammalian CNS}

Information in the brain is transmitted at synapses, which are highly
sophisticated contact zones between a sending and a receiving nerve
cell. They have a typical asymmetric structure where the sending,
presynaptic part is specialized for the secretion of neurotransmitters
and other signaling molecules while the receiving, postsynaptic part
is composed of complex signal transduction machinery.

In the developing human embryo, cell recognition mechanisms with high
resolution generate an ordered network of some $10^{15}$ synapses,
linking about $10^{12}$ nerve cells. The extraordinary specificity
of synaptic connections in the adult brain is generated in several
consecutive steps. Initially, immature nerve cells migrate to their
final location in the brain. There, they form processes, so called
axons. Axons grow, often over quite long distances, into the target
region that the corresponding nerve cell is supposed to hook up with.
Once arrived in the target area, an axon selects its target cell from
a large number of possible candidates. Next, a synapse is formed at
the initial site of contact between axon and target cell. For this
purpose, specialized proteins are recruited to the synaptic contact
zone. Newly formed synapses are then stabilized and modulated, depending
on their use. These processes result in finely tuned networks of nerve
cells that mediate all brain functions, ranging from simple movements
to complex cognitive or emotional behaviour.

Song et al. (1999) have studied the biochemical characteristics and
cellular localization of neuroligin-1, which is a member of a brain-specific
family of cell adhesion proteins. They discovered that neuroligin-1
is specifically localized to synaptic junctions, making it the first
known synaptic cell adhesion molecule. Using morphological methods
with very high resolution Brose (1999) demonstrated that neuroligin-1
resides in postsynaptic membranes, its extracellular tail reaching
into the cleft that separates postsynaptic nerve cells from the presynaptic
axon terminal. Interestingly, the extracellular part of neuroligin-1
binds to another group of cell adhesion molecules, the $\beta$-neurexins.
Based on their findings, Brose and colleagues suggest a novel molecular
model of synapse formation in the brain. Postsynaptic neuroligin interacts
with presynaptic $\beta$-neurexin to form a transsynaptic cell-adhesion
complex at a developing synapse. Once the junction is formed, neuroligins
and $\beta$-neurexins initiate well-characterized intracellular protein-protein-interaction
cascades. These lead to the recruitment of proteins of the transmitter
release machinery on the presynaptic side and of signal transduction
proteins on the postsynaptic side. The resulting transsynaptic link
could also function in retrograde and anterograde signalling of mature
synapses.

\section{Interneuronal entanglement}

Ultrastructural studies of excitatory synapses have revealed an electron-dense
thickening in the postsynaptic membrane known as the postsynaptic
density (PSD). The PSD has been proposed to be a protein lattice that
localizes and organizes the various receptors, ion channels, kinases,
phosphatases and signaling molecules at the synapse (Fanning and Anderson,
1999). Studies from many laboratories over the past ten years have
identified various novel proteins that make up the PSD. Many of these
proteins contain PDZ domains, short sequences named after the proteins
in which these sequence motifs were originally identified (PSD-95,
Discs-large, Zona occludens-1). PDZ domains are protein-protein interaction
motifs that bind to short aminoacid sequences at the C-termini of
membrane proteins. These PDZ domain-containing proteins have been
shown to bind many types of synaptic proteins, including all three
classes of ionotropic glutamate receptors, and seem to link these
proteins together to organize and regulate the complex lattice of
proteins at the excitatory synapse.

CASK is a presynaptic protein that binds to the cell surface protein
$\beta$-neurexin (Hata et al., 1996), while the third PDZ domain
of the postsynaptic protein PSD-95 has been demonstrated to bind to
the C-terminal tail of neuroligins (Irie et al., 1997). CASK and PSD-95
stabilize synaptic structure by mediating interactions with cell adhesion
molecules $\beta$-neurexin (presynaptic) and neuroligin-1 (postsynaptic)
or by indirectly linking synaptic proteins to the cytoskeleton through
the actin binding protein 4.1 (presynaptic) or the microtubule-binding
protein CRIPT (postsynaptic).

Direct binding of neurexins to neuroligins has been demonstrated to
promote cell-cell interactions, leading to the suggestion that adhesive
interactions mediated by PDZ proteins might promote assembly or stabilization
of synaptic structure (Missler and Südhof, 1998). The CASK PDZ domain
has also been shown to bind to syndecans, which are cell surface proteoglycans
implicated in extracellular matrix attachment and growth factor signaling
(Cohen et al., 1998; Hsueh et al., 1998). Both, neuroligins and $\beta$-neurexins
are the cores of well-characterized intracellular protein-protein-interaction
cascades. These link neuroligins to components of the postsynaptic
signal transduction machinery and $\beta$-neurexins to the presynaptic
transmitter secretion apparatus. The $\beta$-neurexin/neuroligin-1
junction provides an interesting and simple mechanism for retrograde
signalling during learning-dependent changes in synaptic connectivity.
Indeed, it allows for direct signalling between the postsynaptic nerve
cell and the presynaptic transmitter secretion machinery. Neurophysiologists
and cognitive neurobiologists have postulated such retrograde signalling
as a functional prerequisite for learning processes in the brain.

The organization of membrane domains might also be mediated by the
ability of many of these multidomain proteins to promote direct or
indirect linkage to cytoskeleton. CASK is tethered to the cortical
cytoskeleton by the actin/spectrin-binding protein 4.1 (Cohen et al.,
1998; Lue et al., 1995). Interestingly, this interaction is mediated
by a conserved module in protein 4.1 know as a FERM domain, which
is known to link other proteins to the plasma membrane (Chishti et
al., 1998). The third PDZ domain of PSD-95 has been demonstrated to
bind to the protein CRIPT, which can recruit PSD-95 to cellular microtubules
in a heterologous cell assay (Niethammer et al., 1998). Linkage of
these scaffolding proteins to the cytoskeleton might help to stabilize
their associated transmembrane proteins within discrete plasma membrane
domains.

PDZ-domain-containing proteins create macromolecular signaling complexes
at both the pre- and postsynaptic junctions, thus defining the active
zone and the postsynaptic density (PSD). Cell-adhesion molecules such
as $\beta$-neurexin and neuroligin seem to keep CASK and PSD-95,
respectively, in register at the pre- and postsynaptic plasma membranes.
CASK interacts with two PDZ-domain-containing proteins, Veli and Mint1.
The latter binds to the cytoplasmic domain of N-type Ca$^{2+}$ channels
possibly ensuring close proximity between the calcium gates and the
exocytotic machinery. Two novel PDZ-domain-containing components of
the presynaptic cytoskeletal matrix are Piccolo and Rim, which might
be involved in mobilization of synaptic vesicles (SVs) from the reserve
to the release-ready pool and in the localization of components of
the endo- and exocytotic machinery. At the postsynaptic plasma membrane,
PDZ-domain-containing proteins from the PSD-95 family are thought
to play a primary role in tethering different glutamate receptors
to the postsynaptic cytoskeletal matrix. CRIPT provides a link to
microtubules, and cortactin a link to actin. PDZ-domain-containing
proteins such as SAP90/PSD-95 and GRIP are also likely to couple glutamate
receptors to the Ras signalling at the postsynaptic site.

We have considered the molecular organization of the synapse to reveal
the molecular connection between the two neuronal cytoskeletons. It
is not surprise that the intrasynaptic $\beta$-neurexin/neuroligin-1
adhesion that is central for synapse formation, not only organizes
the pre- and postsynaptic architecture but also could mediate interneuronal
entanglement. The entangled cytoskeletons then could act as a whole.

\section{The $\beta$-neurexin/neuroligin-1 junction and environmental decoherence}

The intraneuronal proteins can be shielded by the actin meshwork that
order the intraneuronal water molecules, but what about the synaptic
$\beta$-neurexin/neuroligin-1 link? Can it be shielded against environmental
interfering? After all the living cells extract negentropy from their
microenvironment so it is supposed that the order inside the cell
is overcompensated with chaos outside it. Indeed the synaptic $\beta$-neurexin/neuroligin-1
adhesion could mediate interneuronal entanglement because it is surrounded
by synaptic cleft matrix molecules and could be permanently shielded.
Such possible mechanism is shielding by glycosaminoglycans (GAGs),
which interconnect the two neural membranes, or polysaccharides and
intrasynaptic proteoglycans molecules. The chemical synapses are relatively
young devices developed in the evolution of the neural system, so
they should not be considered as extracellular space, rather be considered
as insulated compartment via glial muff, whose function is to ensure
interneuronal communication. Thus the synapse is not as extracellular
microenvironment, but a specialized intercellular device.

\begin{figure}[htp]
\includegraphics[width=110mm]{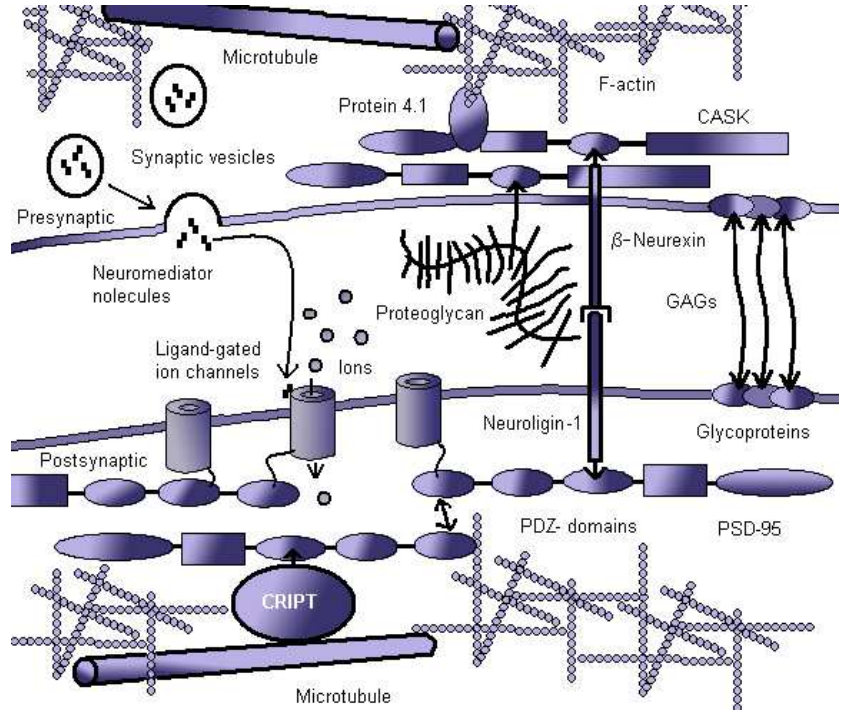}
\centering
\caption{The $\beta$-neurexin/neuroligin-1 adhesion can influence
the cytoskeletons of the two neurons. The quantum coherence between
neurons is mediated by $\beta$-neurexin/neuroligin-1 adhesion which
can be shielded by glycosaminoglycans (GAGs) from decoherence.}
\end{figure}

\subsection{Fucose-galactose bridging of pre- and postsynaptic glycoproteins}

Hsieh-Wilson (2001) has provided experimental data confirming the
special role of fucose in the brain. Fucose is a simple sugar that
is attached to proteins at the synapse and is frequently associated
with other sugar molecules. There is some evidence that fucose is
important for modulating the transmission of signals between two or
more nerve cells. For example, fucose is highly concentrated at the
synapse, and repeated nerve-cell firing increases the levels still
further. Thus fucose may be involved in learning and memory because
disrupting a critical fucose-containing linkage causes amnesia in
lab rats. Fucose is often linked to another sugar called galactose.
The linkage is created when hydroxyl groups on the two sugars combine
and expel a water molecule. Rats given 2-deoxygalactose (which is
identical to galactose in all respects except that it lacks the critical
hydroxyl group) cannot form this linkage, and develop amnesia because
they cannot form the essential fucose-galactose linkage. In another
study, rats treated with 2-deoxygalactose were unable to maintain
long-term potentiation (LTP), which is a widely used model for learning
and memory. Taken together, these experiments strongly suggest that
fucose-containing molecules at the synapse may play an important role
in learning and memory.

Hsieh-Wilson and colleagues have developed a model that may explain
the role of fucose at the synapse. The fucose attached to a protein
on the presynaptic membrane can bind to another protein located at
the postsynaptic membrane. This stimulates the postsynaptic neuron
to make more of the fucose-binding protein, enhancing the cell's sensitivity
to fucose and strengthening the connection. In the quantum brain model
proposed here the polysaccharides interconnecting the pre- and postsynaptic
glycoproteins can order the water molecules in the vicinity of the
$\beta$-neurexin/neuroligin-1 complexes, so that interneuronal polysaccharide
bridges can create ordered microenvironment necessary for $\beta$-neurexin/neuroligin-1
function to entangle the two pre- and postsynaptic cytoskeletons.

\subsection{Synaptic dystrophin glycoprotein complexes}

Localization studies have determined a neuronal distribution of the
dystrophin isoform Dp427, being associated with the postsynaptic density.
Three full-length dystrophin isoforms have been established, resulting
from different promoters, differing only in their N-terminal makeup
and their cellular location (Culligan and Ohlendieck, 2002). The backbone
of the brain dystrophin-glycoprotein complex (DGC) is the transmembrane
link generated by the presence of $\alpha$- and $\beta$-dystroglycan
(Culligan et al., 2001). These proteins act to form an integral plasmalemmal
linkage, localizing dystrophin to the subplasmalemmal region. A proline-rich
region at the extreme C-terminus of $\beta$-dystroglycan mediates
this interaction with dystrophin and cross-linking of brain $\beta$-dystroglycan
results in the stabilization of a high-molecular mass complex (Culligan
et al., 1998). The extracellular matrix component $\alpha$-dystroglycan,
a 156 kD heavily glycosylated protein interacts with the N-terminal
region of $\beta$-dystroglycan. The polymorphic cell-surface proteins,
$\alpha$- and $\beta$-neurexins, have been demonstrated as binding
partners for neuronally expressed dystrophin through an interaction
with $\alpha$-dystroglycan. The interaction links the neuronal postsynaptic
membrane through $\alpha$-dystroglycan with the presynaptic membrane
through the neurexins, mediating cell aggregation (Sugita et al.,
2001). So the neurexin/DGC interneuronal connections can suffice to
dynamically order water molecules in the vicinity forming a shield
for $\beta$-neurexin/neuroligin-1 adhesions.

\begin{figure}[htp]
\includegraphics[width=85mm]{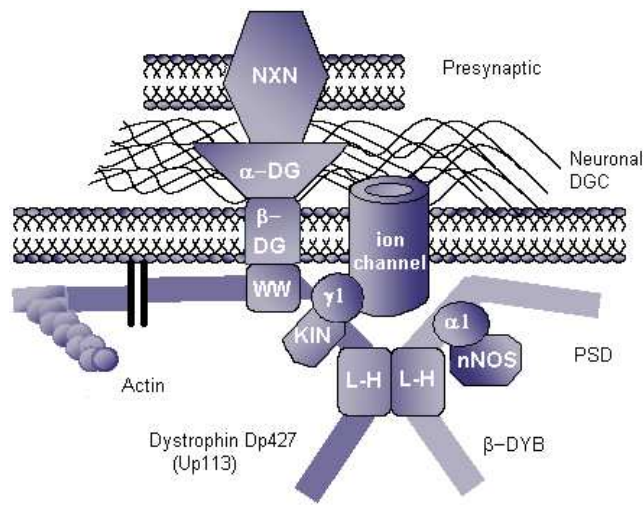}
\centering
\caption{Schematic representation of the established members
of dystrophin-glycoprotein complexes (DGC) and their associations
with other peripheral proteins. A dystrophin or utrophin isoforms
link to the $\alpha/\beta$-dystroglycan (DG) backbone and a dystrobrevin
(DYB) isoform. In neurons, $\alpha$-dystroglycan is associated with
neurexin (NXN), linking the presynaptic membrane to the postsynaptic
density (PSD). Syntrophin isoforms $\alpha1$ and $\gamma1$ recruit
neuronal nitric oxide synthase (nNOS) to the complex, as well as other
nonestablished voltage-gated ion-channels and/or kinases (KIN). Dystrophin
domains: WW motif, common to all brain dystrophin isoforms; L-H motif,
helical leucine heptads which makeup the coiled-coil domain.}
\end{figure}

The role of the DGC in synaptic function is further supported by the
observation that in $\frac{1}{3}$ of the patients affected by Duchenne
muscular dystrophy (DMD) there are brain abnormalities, presented
in the form of a moderate to severe, nonprogressive mental retardation,
are manifest (Mehler, 2000). The abnormality is evident as developmental
cognitive and behavioral abnormalities, including deficits in overall
and verbal IQ, as well as attention deficits and impaired short-term
memory processing (Mehler, 2000; Bresolin et al., 1994).

\subsection{Synaptic sulfated glycosaminoglycans}

Sulfated glycosaminoglycans (GAGs) are linear heteropolysaccharides,
which interconnect the pre- and postsynaptic membranes projecting
negative charged groups, which can help order the positively charged
sodium and potassium ions and repell the chloride ions thus shielding
the $\beta$-neurexin/neuroligin-1 adhesion. GAGs like fucose are
found at the synapse, and are important for proper brain development,
and play a critical role in learning and memory (Hsieh-Wilson, 2001).
It is believed that, like fucose, GAGs are also involved in establishing
connections between nerve cells. 

Whereas fucose is a relatively simple sugar, GAGs are complex polymers,
having a repeating A-B-A-B-A- structure composed of alternating sugar
units. There are several different kinds of GAGs found in nature,
and each GAG is characterized by different sugar units. For example,
chondroitin sulfate is composed of alternating D-glucuronic acid and
N-acetylgalactosamine units.

In the brain D-glucuronic acid may be chemically modified with sulfate
(OSO$_{3}^{-}$) groups at either or both of the 2- and 3- positions.
Every sugar monomer in the GAG molecule is supposed to be slight axially
rotated in respect with the previous one. The sulfate groups itself
could project in different directions thus contributing a bulk of
negative charges in the vicinity ordering water molecules and positive
ions. If the GAGs connecting the two neuronal membranes have proper
space localization they can permanently insulate the $\beta$-neurexin/neuroligin-1
adhesion.

\subsection{Synaptic proteoglycans}

The chondroitin/keratan sulphate proteoglycans of the nervous tissue
may direct the axonal migration. On the other hand, heparan sulphate
chains enhance neurite outgrowth, and they also affect the polarity.
Extracellular chondroitin sulfate proteoglycans seem to decrease cell-cell
and cell-matrix interactions, allowing the cells to round-up, divide,
differentiate and migrate in the tissue. The ability of heparan sulphates
to bind growth factors are possibly important during the growth of
differentiation of nervous cells. Except for their crucial involvement
in the development of the neural system architecture the proteoglycans
stabilize the synaptic structure and fill the synaptic cleft. Here
will be paid attention to two neural proteoglycans: CAT-301 and phosphacan.

The CAT-301 proteoglycan is a developmentally regulated, high molecular
chondroitin sulphate proteoglycan found on the extracellular surface
of mammalian neurons. It is expressed late in development and although
no definitive role has been identified for CAT-301, it is believed
to have a role in the stabilisation of synaptic structure. The name
derives from the name of the monoclonal antibody originally used to
identify it. Disruption of the normal patterns of neuronal activity
during the critical early postnatal period by physical or biochemical
means results in a large and irreversible reduction in levels of CAT-301.
Similar intervention in mature animals has no effect.

Phosphacan (previously designated 3F8 or 6B4) is a chondroitin sulphate
proteoglycan that binds to neurons and neural cell-adhesion molecules
(Maurel et al., 1994; Maeda et al., 1995; Garwood et al., 1999). Cloning
of this proteoglycan showed that it has a high homology with receptor-type
protein tyrosine phosphatase, formed by alternative splicing (Maurel
et al., 1994; Sakurai et al., 1996). It binds with a high affinity
to nervous tissue adhesion molecules Ng-CAM and N-CAM, but not laminin,
fibronectin, or collagens. Tyrosine phosphatases function together
with tyrosine kinases regulating protein phosphorylation, and they
can mediate their actions through signal transduction system of the
cell.

\section{Quantum teleportation between neurons}

In the QED-Cavity model of microtubules Mavromatos et al. (2002) show
that intraneuronal dissipationless energy transfer and quantum teleportation
of coherent quantum states are in principle possible. In the neuron
this is achieved between microtubules entangled through MAPs. The
$\beta$-neurexin/neuroligin-1 entanglement could allow such teleportation
to occur between cortical neurons. The entanglement can be used for
quantum transfer of tubulin states between neurons; the state of the
recipient microtubule then could affect specific intraneuronal processes. 

The consciousness is known to be product of the cerebral cortex activity.
We can realize or experience something only if there is proper stimulation
of certain areas within the brain cortex. In the $\beta$-neurexin/neuroligin-1
quantum model of consciousness the interneuronal entanglement is supposed
to occur only between cortical neurons. However arises the question
why quantum coherence cannot be achieved between subcortical or spinal
neurons considering that $\beta$-neurexin/neuroligin-1 link is widely
presented in the CNS? I suppose that the answer should come from studying
the unique molecular synapse structure between the cortical neurons.

\section{The hands of consciousness}

If our conscious mind is in the quantum coherent cytoskeleton then
it should have some power to influence the synaptic activity using
its free will in a uniform way. This is so because everyone can immediately
move his arm, leg etc. or say something. However because nobody can
commit to memory at once a poem this means that the conscious thought
acts much slower on synaptic plasticity. Such kind of arguments can
show us, which brain activities are immediately connected with our
free will, or with the possibility of our mind to collapse the wave
function (motion) and which brain activities are only influenced by
our internal thoughts (memory storage, motor protein dynamics and
synaptic plasticity etc.). Of course, we are not consciously aware
how exactly both types of activities are acted upon by the cytoskeleton
- our consciousness just does it, it influences directly some of the
intracellular processes and has diverse effects in the cortical neurons. 

Following our own experience we can say that our consciousness is
causally effective and its actions depend on our will. If microtubules
just do quantum computing how this could affect the immediate neuromediator
release. Stuart Hameroff supposes that microtubules control the axonal
hillock potential but do not provide any concrete mechanism for that.
It is quite dubious that such exists, because the axonal hillock potential
depends on voltage gated ion channels that do open by changes in the
membrane potential.

Somehow surprisingly the model including the $\beta$-neurexin/neuroligin-1
entanglement not also answers how interneuronal quantum coherence
can be achieved, but also gives answer how the synaptic vesicle release
can be acted upon. The $\beta$-neurexins are essential ligands for
synaptotagmin-1, a protein that docks the synaptic vesicles to the
presynaptic membrane and acts as a Ca$^{2+}$-sensor. Thus the conformational
states of $\beta$-neurexin either directly via synaptotagmin-1 or
indirectly via CASK and Mint-1 could control the exocytosis.

The basis of our new understanding of consciousness is that it is
a fundamental feature of reality and is something dynamic describable
by complex quantum wave born into existence by a conglomerate of entangled
proteins: tubulins, MAP-2, $\beta$-neurexin, neuroligin, CASK, CRIPT,
PSD-95, protein 4.1 etc. In this new model every protein species has
its unique intraneuronal function. Thus a fully functional body for
the mind is built up.

All intraneuronal processes (synaptic plasticity, memory) are influenced
by the protein body of the consciousness: the motor proteins are moving
over the microtubules, the $\beta$-neurexin/neuroligin-1 adhesion
is the core of a new formed synapse, the neuromediator receptors are
anchored to and organized by the cytoskeleton, the synapsins are docking
the synaptic vesicles to the cytoskeleton, the scaffold proteins drive
exocytosis etc. Some of this molecules (kinesin, dynein, neuromediator
receptors), different types of vesicles, actin filaments, enzymes
etc. are possibly not in coherence with our conscious quantum state,
so our consciousness is influencing them not so easy, not so fast,
and not by will.

\section{Quantum tunnelling and neuromediator release}

The model proposed by Friedrich Beck and Sir John Eccles (1992) introduces
a quantum element into the functioning of the brain through the mechanism
of exocytosis, the process by which neurotransmitter molecules contained
in synaptic vesicles are expelled into the synaptic cleft from the
presynaptic terminal. The arrival of a nerve impulse at an axon terminus
does not invariably induce the waiting vesicles to spill their neurotransmitter
content into the synapse, as was once thought. Beck argues that empirical
work suggests a quantum explanation for the observed probabilistic
release, and offers supporting evidence for a trigger model of synaptic
action. The proposed model is realized in terms of electron transfer
processes mediating conformational change in the presynaptic membrane
via tunneling.

Frederick Beck cites the {}``non-causal logic of quantum mechanics,
characterized by the famous von Neumann state reduction'' as the
reason a quantum mechanism might be relevant to the explanation of
consciousness and suggests that probabilistic release at the synaptic
cleft may be the point at which quantum logic enters into the determination
of brain function in an explanatorily non-trivial manner. He postulates
that global activation patterns resulting from non-linear feedback
within the neural net might enhance weak signals through stochastic
resonance, a process by which inherently weak signals can be discerned
even when their amplitudes lie below the level of the ambient background
noise. This might allow sufficient leverage to amplify the role of
the quantum processes governing synaptic transmission to a level that
could be causally efficacious in determining consciousness.

According to Beck the synaptic exocytosis of neurotransmitters is
the key regulator in the neuronal network of the neocortex. This is
achieved by filtering incoming nerve impulses according to the excitatory
or inhibitory status of the synapses. Findings by Jack et al. (1981)
inevitably imply an activation barrier, which hinders vesicular docking,
opening, and releasing of transmitter molecules at the presynaptic
membrane upon excitation by an incoming nerve impulse. Redman (1990)
demonstrated in single hippocampal pyramidal cells that the process
of exocytosis occurs only with probability generally much smaller
than one upon each incoming impulse.

There are principally two ways by which the barrier can be surpassed
after excitation of the presynaptic neuron: the classical over-the-barrier
thermal activation and quantum through-the-barrier tunneling. The
characteristic difference between the two mechanisms is the strong
temperature dependence of the former, while the latter is independent
of temperature, and only depends on the energies and barrier characteristics
involved.

\subsection{Thermal activation}

This leads, according to Arrhenius' law, to a transfer rate, $k$,
of 

\begin{equation}
k\simeq V_{C}\exp\left[-\frac{E_{A}}{k_{B}T}\right]\end{equation}

where $V_{C}$ stands for the coupling across the barrier, and $E_{A}$
denotes the activation barrier.

\subsection{Pure quantum tunnelling}

In this case the transfer rate, $k$, is determined in a semiclassical
approximation by

\begin{equation}
k\approx\omega_{0}\exp\left[-2\int_{a}^{b}\frac{\sqrt{2m(V(q)-E)}}{\hbar}dq\right]\end{equation}

where $m$ is the mass of the tunneled quantum particle, $E$ is the
energy of the quantum particle, $V(q)$ is the collective potential
for the motion of the quasiparticle that triggers exocytosis, $\omega_{0}=\frac{E_{0}}{\hbar}$
is the number of attempts that the particle undertakes to reach the
barrier, and $E_{0}$ is the energy of the quasi-bound tunneling state
that is the quantum mechanical zero-point energy of a particle of
mass $m$ localized over a distance of $\Delta q$. The quantum trigger
model for exocytosis developed by Beck and Eccles (1992) is based
on the idea of pure quantum tunneling. The reason for this choice
lies in the fact that thermal activation is a broadly uncontrolled
process, depending mainly on the temperature of the surroundings,
while quantum tunneling can be fine-tuned in a rather stable manner
by adjusting the energy $E_{0}$ of the quasi-bound state or, equivalently,
by regulating the barrier height (the role of the action potential).

\begin{figure}[htp]
\includegraphics[width=70mm]{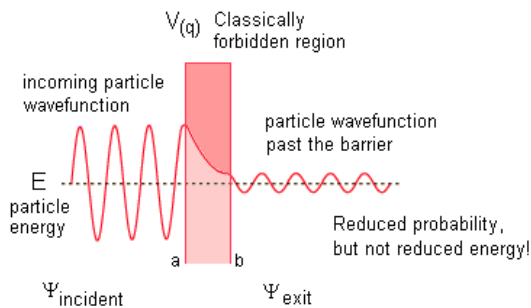}
\centering
\caption{Schematic representation of the pure quantum tunnelling
process. According to classical physics, a particle of energy $E$
less than the height $V(q)$ of a barrier could not penetrate - the
region inside the barrier is classically forbidden. But the wavefunction
$\Psi$ associated with a free particle must be continuous at the
barrier and will show an exponential decay inside the barrier. The
wavefunction must also be continuous on the far side of the barrier,
so there is a finite probability that the particle will tunnel through
the barrier. Legend: $a$ and $b$ are the classical turning points
of the motion inside and outside the barrier.}
\end{figure}

A careful study of the energies involved showed that quantum tunneling
remains safe from thermal interference only if the tunneling process
is of the type of a molecular transition, and not a quantum motion
in the macromolecular exocytosis mechanism as a whole. In further
work Beck attributed the molecular tunneling to the electron transfer
mechanism in biomolecules (Beck, 1996; Beck and Eccles, 2003). The
current experimental data however seem to support an exocytosis model
in which the transmission rate is dependent on temperature. However
if the tunnelling in exocytosis is multidimensional as shown in certain
enzymes then it could use the thermal fluctuations - so called vibrationally
assisted tunneling.

\subsection{Mixed quantum tunneling}

One of the most outstanding differences between one-dimensional and
multidimensional tunnelling is the possibility for mixed tunnelling
in the multidimensional case. The mixed tunnelling is such a tunnelling
that classical motion is allowed in one or more directions in the
multidimensional space. In contrast in the pure tunnelling classical
motion is not allowed in any direction. The vibrationally assisted
tunnelling in enzymatic action described further in the text is a
kind of mixed tunnelling.

It was found that there are three kinds of vibrational modes with
respect to the effects of the excitation on tunnelling; 1) those which
do not affect the tunnelling, 2) those which promote the tunnelling,
and 3) those which suppress the tunnelling. If there is no coupling
between the tunnelling coordinate and the coordinate transversal to
it, the vibrational excitation in the latter does not affect energy
splitting. This corresponds to the first type. When there is a coupling
between the two coordinates, it is natural to expect from the analogy
with the one-dimensional case that the vibrational excitations promote
the tunnelling. However, the experimental findings in tropolone molecules
clearly show that the real proton tunnelling is not so simple. This
fact nicely exemplifies the complexity of multidimensional tunnelling.

Takada and Nakamura (1995) show that depending on the topography of
potential energy surface (PES) vibrational excitation either promote
or suppress the tunnelling and that the mixed tunnelling plays an
essential role in suppression and oscillation of energy splitting
against vibrational excitation. The general Wentzel-Kramers-Brillouin
(WKB) theory of multi-dimensional tunnelling developed by Takada and
Nakamura (1994, 1995) provides us with a clear conceptual understanding
of the multidimensionality. The theory was formulated by solving the
following basic problems: (i) construction of the semiclassical eigenfunction
in classically allowed region according to the Maslov theory, (ii)
its connection to the wave function in the classically inaccessible
region, and (iii) propagation of the latter into the deep tunnelling
region. 

It became clear that there exist two distinct tunnelling regions:
$C$-region where action is complex and $I$-region where action is
pure imaginary. Tunnelling in these regions is qualitatively quite
different from each other; in the $I$-region the tunnelling path
can be defined by a certain classical trajectory on the inverted potential,
while in the $C$-region there is no unique path and the Huygens type
wave propagation should be applied.

\begin{figure}[htp]
\includegraphics[width=60mm]{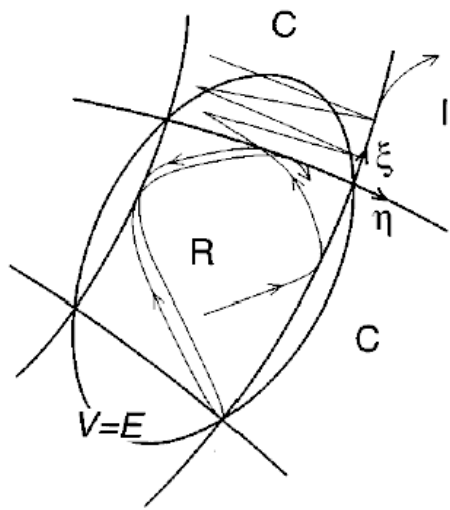}
\centering
\caption{Schematic drawing of the physical picture of tunneling
obtained by the WKB theory. The classical trajectories comprising
a quantum eigenstate are confined within the distorted rectangular
region (called $R$-region), although the much wider region is energetically
allowed (see the oval region bounded by $V=E$). The tunnelling proceeds
first to the $C$-region where the motion in $\xi$-direction is non-classical
(tunnelling), while the motion in $\eta$-direction is still classical.
This type of tunnelling is \emph{mixed tunnelling}. At the boundary
between the $C$-region and $I$-region, part of the tunnelling wave
enters into the $I$-region where no classical motion is allowed in
any direction. This conventional type of tunnelling is called \emph{pure
tunneling}.}
\end{figure}

In the synaptic vesicle release there are involved a number of proteins
(SNAP-25, synaptotagmin-1, synaptobrevin-1, $\beta$-neurexin, syntaxin-1)
and because there is an energy barrier their function can be compared
with the enzyme catalytic action. According to Hameroff (1998) the
London quantum forces set the pattern for protein dynamics. A year
later Basran et al. (1999) have found evidence for vibration driven
extreme tunneling for enzymatic proton transfer.

At the beginning of 21$^{\text{th}}$ century the Haldane's notion
of {}``imperfect key'' about the biological catalysis in classical
over-the-barrier manner is questioned. Sutcliffe and Scrutton (2000a)
underline that matter is usually treated as a particle. However it
can also be treated as a wave (wave-particle duality). These wavelike
properties, which move our conceptual framework into the realm of
quantum mechanics, enable matter to pass through regions that would
be inaccessible if it were as a particle. In the quantum world, the
pathway from reactants to products might not need to pass over the
barrier but pass through the barrier by quantum tunnelling. Quantum
tunnelling is more pronounced for light particles (e.g. electrons),
because the wavelength of a particle is inversely proportional to
the square root of the mass of the particle. Electrons can be tunnelled
for distance of about 3 nm. Protium can tunnel over a distance of
0.058 nm with the same probability as an electron tunnelling over
2.5 nm. The isotopes of hydrogen - deuterium and tritium have increased
mass and tunnel with the same probability over 0.041 nm and 0.034
nm. Klinmann and coworkers were the first to obtain experimental evidence
consistent with hydrogen tunnelling in an enzyme-catalyzed reaction
on the basis of deviation in kinetic isotope effect from that expected
for classical behaviour. Since their proposal of hydrogen tunnelling
at physiological temperatures in yeast alcohol dehydrogenase (Cha
et al., 1989), they have also demonstrated similar effects in bovine
serum amine oxidase (Gant and Klinman, 1989), monoamine oxidase (Johnsson
et al., 1994) and glucose oxidase (Kohen et al., 1997). Tunnelling
in these systems was described in terms of static barrier depictions. 

The pure quantum tunnelling reactions are temperature independent,
because thermal activation of the substrate is not required to ascend
the potential energy surface. However the rate of C-H and C-D cleavage
by methylamine dehydrogenase was found to be strongly dependent on
temperature, indicating that thermal activation or {}``breathing''
of the protein molecule is required for catalysis. Moreover, the temperature
dependence of the reaction is independent of isotope, reinforcing
the idea that protein (and not substrate) dynamics drive the reaction
and that tunnelling is from the ground state. Good evidence is now
available for vibrationally-assisted (mixed) tunnelling (Bruno and
Bialek, 1992; Basran et al., 1999) from studies of the effects of
pressure on deuterium isotope effects in yeast alcohol dehydrogenase
(Northrop and Cho, 2000). Combining the experimental evidence, the
argument for vibrationally-assisted tunnelling is now compelling.
The dynamic fluctuations in the protein molecule are likely to compress
transiently the width of the potential energy barrier and equalize
the vibrational energy levels on the reactant and product site of
the barrier (Sutcliffe and Scrutton, 2000b; Scrutton et al., 1999;
Kohen and Klinman, 1999). Compression of the barrier reduces the tunnelling
distance (thus increasing the probability of the transfer), and equalization
of vibrational energy states is a prerequisite for tunnelling to proceed.
Following transfer to the product side of the barrier, relaxation
from the geometry required for tunnelling traps the hydrogen nucleus
by preventing quantum leakage to the reactant side of the barrier. 

Catalysis is driven by quantum fluctuations that affect the protein
conformations. We could therefore generalize that every protein driven
process (transport, muscle contraction, exocytosis) could be referred
to as a catalysed process and thus quantum in nature. The most important
here is to note that the pure quantum tunnelling is temperature independent.
In contrast it seems feasible that proteins have evolved mechanisms
to utilize the thermal energy via vibrationally assisted tunnelling.
Therefore one might expect exocytosis and neuromediator release to
be driven by vibrationally-assisted quantum tunneling and not by pure
tunneling as suggested by Beck and Eccles.

\end{document}